\documentstyle[epsfig]{aipproc}

\begin{document}
\title{The Charm of the Proton and the $\Lambda _c^{+}$  Production
\thanks{This work was partially supported by Centro Latino Americano de 
F\'{\i}sica (CLAF).}}

\author{J. dos Anjos, G. Herrera\thanks{Permanent address: CINVESTAV, Apdo 
Postal 14-740 M\'exico DF,M\'exico}, J. Magnin and F.R.A. Sim\~ao}
\address{CBPF, Rua Dr. Xavier Sigaud 150, CEP 22290-180 - Urca RJ,
Rio de Janeiro, Brazil}

\maketitle
\begin{abstract}
We propose a two component model for charmed baryon production
in $pp$ collisions consisting of the conventional parton fusion mechanism 
and fragmentation plus quarks recombination in which a $ud$ valence 
diquark from the proton recombines with a $c$-sea quark to produce a 
$\Lambda_c^+$.
Our two-component model is compared with the intrinsic charm
two-component model and experimental data.
\end{abstract}

\section*{Introduction}

The production mechanism of hadrons containing heavy quarks is not well
understood. Although the fusion reactions $gg\rightarrow Q\bar{Q}$ and
$q\overline{q}\rightarrow Q\bar{Q}$ are supposed to be the dominant
processes, they fail to explain important features of heavy quark
hadro-production like the leading particle effects observed in $D^{\pm}$
produced in $\pi^- p$ collisions \cite{e791}, $\Lambda_c^+$ production
in $pp$ interactions \cite{bari} \cite{isr} and in others
baryons containing heavy quarks \cite{otros}, the $J/\Psi$ cross section
at large $x_F$ observed in $\pi p$ collisions \cite{na3}, etc.\\

The above mentioned effects have been explained using a two-component
model \cite{vogt} consisting of the parton fusion mechanism
calculable in perturbative QCD plus the coalescence of intrinsic charm 
\cite{bhps-plb}.\\

In hadron-hadron collisions the recombination of valence spectator quarks 
with $c$-quarks present in the sea of the initial hadron is a 
possible mechanism for charmed hadron production. Here we explore that 
possibility for the $\Lambda_c^+$'s
production in $pp$ interactions.
We will assume that in addition to the usual parton fusion processes,
a $ud$ diquark recombines with $c$-sea quark both from the incident proton.\\

We compare our results with those of the intrinsic charm
two-component model and the experimental data available.\\

\section*{\label{Dos} $\Lambda_c^{+}$ Production {\it via } Parton Fusion}

In the parton fusion mechanism the $\Lambda _c^{+}$ is produced via the
subprocesses $q\bar{q}(gg) \rightarrow c \bar{c}$ with the
subsequent fragmentation of the $c$ quark.
The inclusive $x_F$ distribution of the $\Lambda _c^{+}$ in $pp$
collisions is given by \cite{vbh-npb}\cite{ram-brod}

\begin{equation}  \label{uno}
\frac{d\sigma^{pf} }{dx_F}=\frac{1}{2} \sqrt{s} \int H_{ab}(x_a,x_b,Q^2)
\frac{1}{E} \frac{D_{\Lambda_c/c} \left( z \right)}{z} dz dp_T^2 dy \: ,
\end{equation}
where
\begin{eqnarray}
H_{ab}(x_a,x_b,Q^2)& = & \Sigma_{a,b} \left( q_a(x_a,Q^2)
\bar{q_b}(x_b,Q^2) \right. \nonumber \\
                   &   & + \left. \bar{q_a}(x_a,Q^2)
q_b(x_b,Q^2) \right) \frac{d \hat{\sigma}}{d \hat{t}} \mid_{q\bar{q}}
\nonumber \\
                   &   & + g_a(x_a,Q^2) g_b(x_b,Q^2)
\frac{d \hat{\sigma}}{d \hat{t}}\mid_{gg}
\end{eqnarray}
\noindent
with $x_a$ and $x_b$ being the parton momentum fractions, $q(x,Q^2)$ and
$g(x,Q^2)$ the quark and gluon distribution in the proton, $E$ the
energy of the produced $c$-quark and $D_{\Lambda_c/c} \left( z \right)$
the fragmentation function. In eq. 1, $p_T ^2$ is the squared 
transverse momentum of the produced $c$-quark, $y$ is the
rapidity of the $\bar {c}$ quark and $z=x_F/x_c$ is the momentum fraction
of the charm quark carried by the $\Lambda _{c}^{+}$. The sum in
eq. 2 runs over $a,b = u,\bar{u},d,\bar{d},s,\bar{s}$.\\

We use the LO results for the elementary cross-sections 
$\frac{d\hat{\sigma}}{d \hat{t}}\mid_{q \bar {q}}$ and 
$\frac{d \hat{\sigma}}{d \hat{t}}\mid_{gg}$ \cite{vbh-npb}.

\begin{equation}
\frac{d \hat{\sigma}}{d \hat{t}}\mid_{q \bar {q}} =
\frac{\pi \alpha _{s}^{2}
 \left( Q^2 \right)}{9 \hat{m}_{c}^{4}} \; \frac{cosh \left( \Delta y
\right) + m_{c}^{2}/ \hat{m}_{c}^{2}}{\left[ 1+cosh \left( \Delta y
\right) \right] ^3}
\end{equation}

\begin{equation}
\frac{d \hat{\sigma}}{d \hat{t}}\mid_{gg}= \frac{\pi \alpha_{s}^{2}
\left( Q^2 \right)}{96 \hat{m}_{c}^{4}} \; \frac{8 cosh
\left( \Delta y \right) -1}{\left[ 1+cosh \left( \Delta y \right)
\right]^3} \: \left[ cosh \left( \Delta y \right)+
\frac{2m_c^2}{\hat{m}_c^2}+\frac{2m_c^4}{\hat{m}_c^4}\right]
\end{equation}
\noindent
where $\Delta y$ is the rapidity gap between the produced $c$ and
$\bar{c}$ quarks and $\hat{m}_c^2=m_c^2+p_T^2$.

In order to be consistent with the LO calculation of the elementary cross
sections, we use the GRV-LO parton distribution functions \cite{gr-zpc},
allowing by a global factor $K \sim 2-3$ in eq. 1 to take into
account NLO contributions \cite{vogt}.\\

We take $m_c=1.5 \:GeV$ for the $c$-quark mass and fix the scale of
the interaction at $Q^2 = 2m_c^2$ \cite{vbh-npb}. Following \cite{vogt},
we use two fragmentation functions to describe the hadronization of
the charm quark;
\begin{equation}
D_{\Lambda_c/c}(z) =  \delta(1-z)
\end{equation}
and the Peterson fragmentation function \cite{peterson}
\begin{equation}
D_{\Lambda_c/c}(z) = \frac{N}{z \left[ 1 - 1/z - 
\epsilon_c/(1-z) \right]^2}
\end{equation}
\noindent
with $\epsilon_c= 0.06$ and the normalization defined by
$\sum _{H} \int D_{H/c}(z) dz = 1$.\\

\section*{$\Lambda _c^{+}$ Production {\it via} Recombination}

The production of leading mesons at low $p_T$ by
recombination of quarks was proposed long time ago \cite{dh-plb}.
The method introduced by Das and Hwa for mesons was
extended by Ranft \cite{ranft-pr} to describe single particle distributions
of leading baryons in $pp$ collisions.\\

In recombination models one assumes that the outgoing hadron is
produced in the beam fragmentation region through the recombination of
the maximun number of valence and the minimun number of sea
quarks coming from the projectile according to
the flavor content of the final hadron. Thus, {\it e.g.} $\Lambda _c^+$'s
produced in $pp$ collisions are formed by the $ud$ valence
diquark and a $c$-quark from the sea of the incident proton.
One ignores other type of contributions involving more
than one sea flavor recombination.\\
The invariant inclusive $x_F$ distribution for leading baryons is given by

\begin{equation}
\frac{2 E}{\sqrt {S}\sigma}\frac{d\sigma^{rec}
}{dx_F}=\int_0^{x_F}\frac{dx_1}{x_1}\frac{dx_2}{x_2}\frac{dx_3
}{x_3}F_3\left( x_1,x_2,x_3\right) R_3\left( x_1,x_2,x_3,x_F\right)
\end{equation}
\noindent
where $x_i$, $i=1,2,3$, is the momentum fraction of the $i^{th}$
quark, $F_3 \left( x_1,x_2,x_3 \right) $ is the three-quark distribution
function in the incident hadron and $R_3\left( x_1,x_2,\right.$
$\left.x_3,x_F\right) $ is the three-quark recombination function. \\
We use a parametrization containing explicitly the single quark
distributions for the three-quark distribution function
\begin{equation}
F_3 \left( x_1,x_2,x_3 \right) = \beta F_{u,val}\left(x_1\right)
F_{d,val}\left(x_2\right)F_{c,sea}\left(x_3\right)
\left(1-x_1-x_2-x_3\right)^{\gamma}
\end{equation}
with $F_{q}\left(x_i\right) = x_iq\left(x_i\right)$ and $F_u$ normalized
to one valence $u$ quark. The parameters $\beta$ and $\gamma$ are
constants fixed by the consistency condition
\begin{eqnarray}
F_{q}\left(x_i\right) & = & \int_0^{1-x_i}dx_j \int_0^{1-x_i-x_j}dx_k \:
F_3 \left( x_1,x_2,x_3 \right), \nonumber \\
                      &   &i,j,k = 1,2,3 \;
\end{eqnarray}
\noindent
for the valence quarks of the incoming proton as in ref. \cite{ranft-pr}.

We use the GRV-LO parametrization for the single quark distributions
in eq. 8. It must be noted that since the
GRV-LO distributions are functions of $x$ and $Q^2$, 
then our $F_3\left( x_1,x_2,x_3 \right)$ also depends on $Q^2$. \\

In contrast with the parton fusion calculation, in which the scale
$Q^2$ of the interaction is fixed at the vertices of the appropriated
Feynman diagrams, in recombination there is not clear way to fix the
value of the parameter $Q^2$, which in this case
is not properly a scale parameter and should be used to give adequately
the content of the recombining quarks in the initial hadron.\\

Since the charm content in the proton sea increases rapidly for $Q^2$
growing from $m_c^2$ to $Q^2$ of the order of some $m_c^2$'s when it
become approximately constant, we take $Q^2 = 4 m_c^2$, a conservative
value, but sufficiently far from the charm threshold in order to avoid a 
highly depressed charm sea which surely
does not represent the real charm content of the proton.
At this value of $Q^2$ we found that the condition of
eq. 9 is fulfilled approximately with $\gamma = -0.1$ and
$\beta = 75$.  We have verified that the recombination cross section
does not change appreciably at higher values of $Q^2$.\\

For the three-quark recombination function for $\Lambda_c^+$ production
we take the simple form \cite{ranft-pr}

\begin{equation} R_3\left( x_u,x_d,x_c\right) =\alpha
\frac{x_ux_dx_c}{x_F^2}\delta \left( x_u+x_d+x_c-x_F\right) \: 
\end{equation} 
\noindent
with $\alpha$ fixed by the condition 
$\int_0^1 dx_F(1/\sigma)d\sigma^{rec}/dx_F = 1$, then $\sigma$ is 
the cross section
for $\Lambda_c^+$'s inclusively produced in $pp$ collisions.
From eqs 7 and 8, the invariant $x_F$ distribution for $\Lambda_c$ is 
\begin{eqnarray}
\frac {2 E}{\sqrt {s} \sigma} \frac{d\sigma_{\Lambda_c^+}^{rec}}{dx_F}
& = & 75 \alpha
\frac{(1-x_F)^{-0.1}}{x_F^2} \int_0^{x_F} dx_1 F_{u,val}(x_1) \nonumber \\
 &   & \times \int_0^{x_F-x_1} dx_2 F_{d,val}(x_2) F_{c,sea}(x_F-x_1-x_2)
\end{eqnarray}
\noindent
where we already integrated over $x_3$. The parameter $\sigma$ will be 
fixed with experimental data.\\

The inclusive production cross section of the $\Lambda_c^{+}$ is
obtained by adding the contribution of recombination
eq. 11 to the QCD processes of eq. 7, then

\begin{equation}
\frac{d\sigma^{tot}}{dx_F} = \frac{d\sigma^{pf} }{dx_F} +
\frac{d\sigma^{rec}}{dx_F}.
\end{equation}

The resulting inclusive $\Lambda_c^{+}$ production cross section
$d\sigma^{tot}/dx_F$ is plotted in fig. 1 using the
two fragmentation function of eqs. 5 and 6 and
compared with experimental data in $pp$ collisions from the ISR
\cite{isr}. As we can see, the shape of the experimental
data is very well described by our model. We use a factor
$\sigma = 0.92 (0.72) \mu bar$ for Peterson (delta)
fragmentation respectively.

In a similar approach  R. Vogt {\it et al.} \cite{vogt} calculated
the $\Lambda _c^{+}$ production in $pp$ and $\pi p$ collisions.
The two component model used by them consists of a parton fusion
mechanism plus coalescence of the intrinsic charm in the proton.
Their results are shown in fig.1. The normalization however 
has been modified to make a proper comparison to our result.\\

\begin{figure}[t]
\begin{center}
\begin{minipage}[t]{2.5in}
\psfig{figure=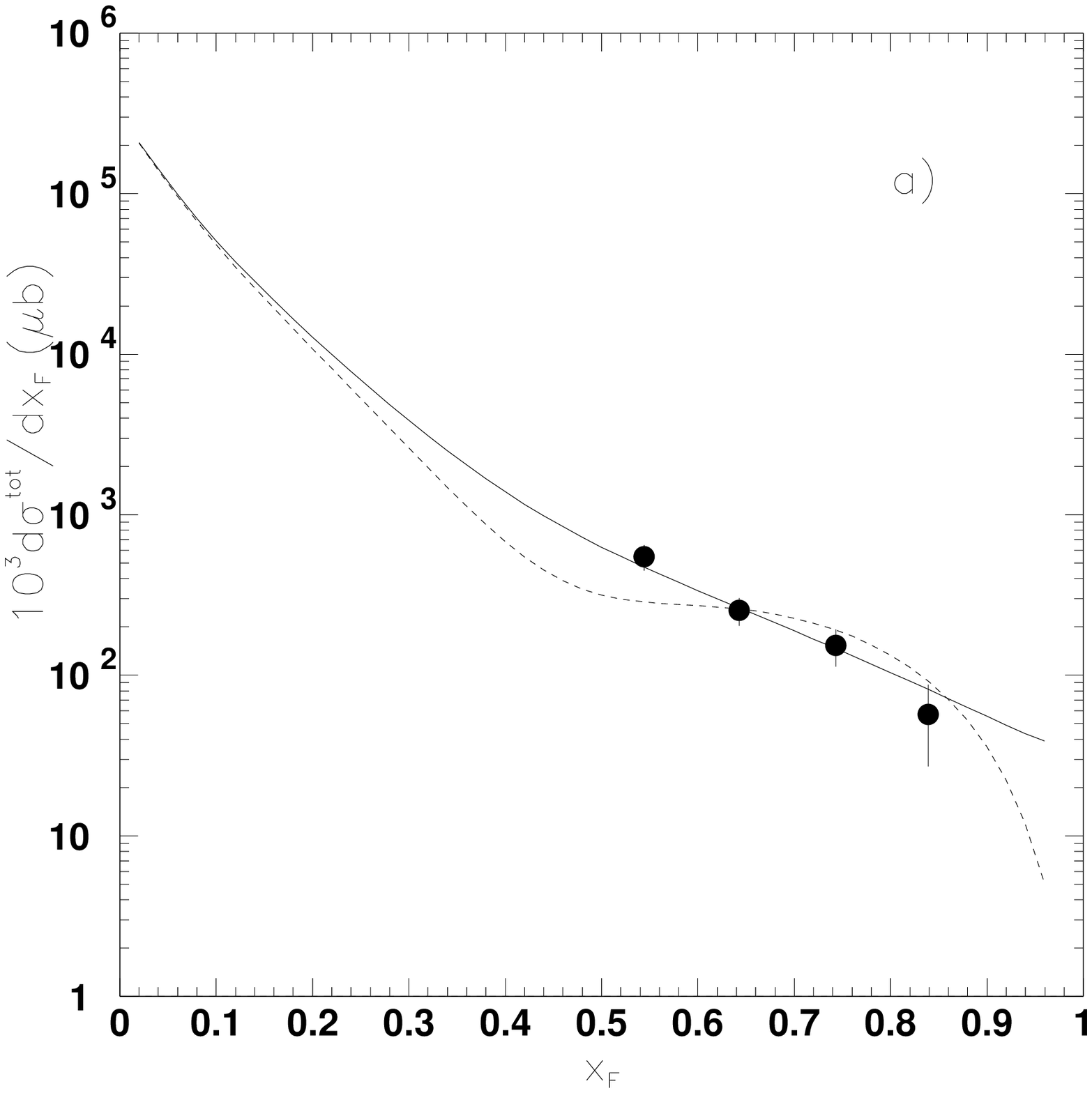,height=2.9in}
\end{minipage}
\begin{minipage}[t]{2.5in}
\psfig{figure=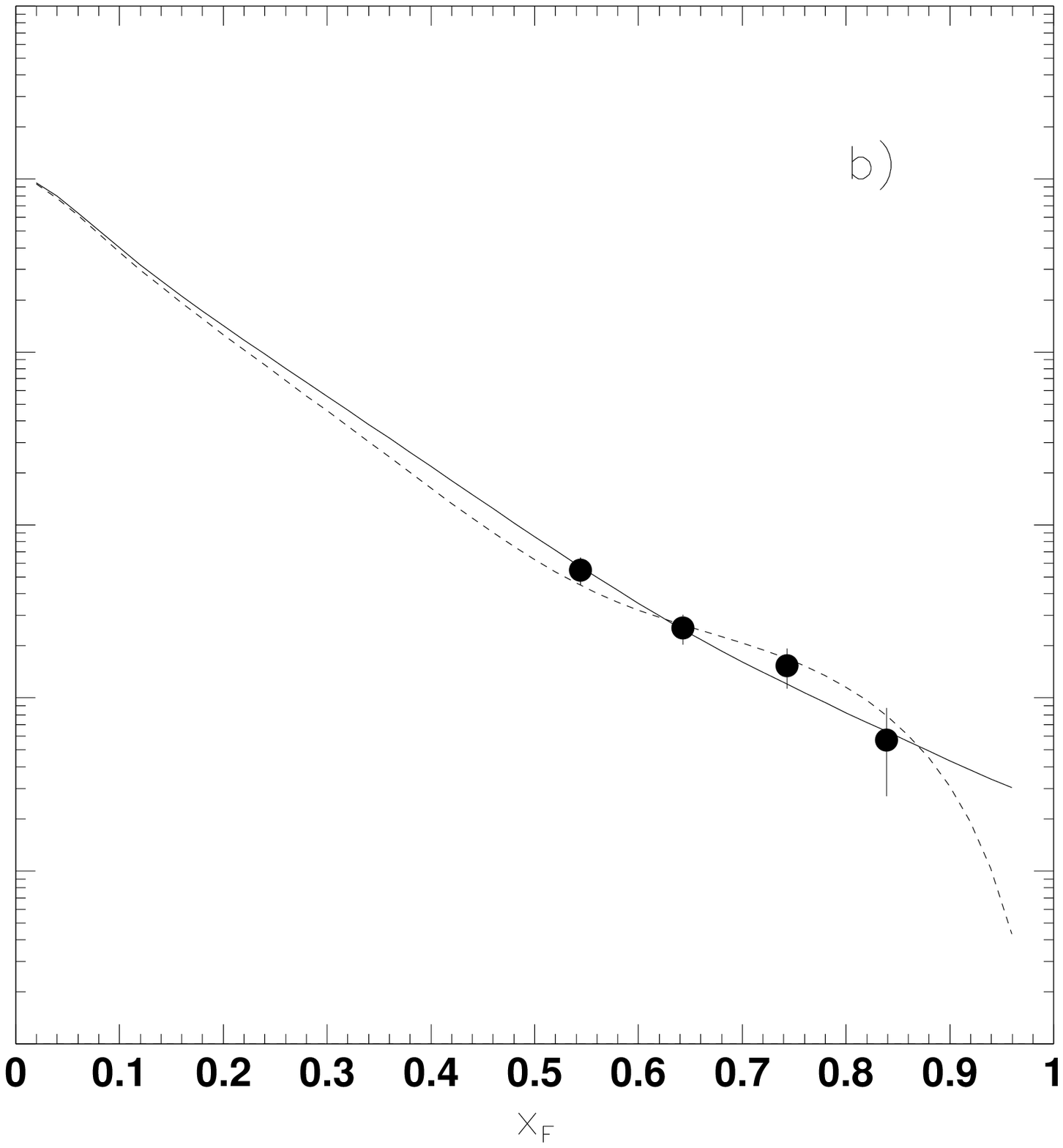,height=2.9in}
\end{minipage}
\end{center}
\caption{
$x_F$ distribution predicted by parton fusion plus recombination
(full line) and parton fusion plus IC coalescence (dashed line) for
Peterson fragmentation (a) and delta fragmentation function (b).
Experimental data (black dots) are taken from ref. 3.}
\end{figure}

\section*{Conclusions} 
We have studied the $\Lambda_c^+$ production in
$pp$ collisions with a two component model.
We show that both the intrinsic charm model and the conventional 
recombination of quarks can describe the 
shape of the $x_F$ distribution for
$\Lambda_c^+$'s produced in $pp$ collisions. 
None of them, however, can
describe the abnormally high normalization of the ISR data quoted in
ref. \cite{isr}.
This discrepancy between theory and experiment does
not exist for charmed meson production, which is
well described both in shape and normalization with the parton
fusion mechanisn plus intrinsic charm coalescence \cite{ram-brod} and
with the conventional recombination as proposed here \cite{nos}.\\
An interesting test to rule out one of the two models would
come from a measurement of the $\Lambda_c$ polarization as proposed
in \cite{herrera}.

\section*{Acknowledgments}

We would like to thank the organizers for financial support to 
participate in the I Simposium Latino Americano de F\'{\i}sica de 
Altas Energ\'{\i}as and for the kind hospitality extended to us during 
the event.


\begin{references}

\bibitem{e791} E791 Collaboration (E.M. Aitala {\it et al.}), Phys.
Lett.  {\bf B 371},157 (1996).

\bibitem{bari}  A.N. Aleev {\it et al.}, Z. Phys. {\bf C 23}, 333 (1984),
G. Bari {\it et al.}, Nuovo Cimento, {\bf 104 A}, 57 (1991).

\bibitem{isr} P. Chauvat {\it et al.} Phys. Lett. {\bf B 199}, 304 (1987).

\bibitem{otros} S.F. Biagi {\it et al.}, Z. Phys. {\bf C 28}, 175 (1985),
G. Bari {\it et al.}, Nuovo Cimento {\bf 104 A}, 787 (1991), R. Werding,
WA89 Collaboration, in proceeding of ICHEP94, Glasgow.

\bibitem{na3} NA3 Collaboration (J. Badier {\it et al.}),
Z. Phys. {\bf C 20}, 101 (1983).

\bibitem{vogt} R. Vogt and S.J. Brodsky, Nucl. Phys. {\bf B 478}, 311 (1996).

\bibitem{bhps-plb}  S.J. Brodsky, P. Hoyer, C. Peterson and N. Sakai, Phys
Lett. {\bf B 93}, 451 (1980) and S.J. Brodsky, C. Peterson and N. Sakai,
Phys. Rev. {\bf D 23}, 2745 (1981).

\bibitem{vbh-npb} J. Babcock, D. Sivers and S. Wolfram,
Phys. Rev. {\bf D 18}, 162 (1978), B.L. Combridge,
Nucl. Phys. {\bf B 151}, 429 (1979), P. Nason, S. Dawson and R.K. Ellis,
Nucl. Phys. {\bf B 327}, 49 (1989), R.K. Ellis, Fermilab-Conf-89/168-T (198
9), I. Inchiliffe, Lectures at the 1989 SLAC Summer Institute, LBL-28468 
(1989).

\bibitem{ram-brod} R. Vogt, S.J. Brodsky and P. Hoyer,
Nucl. Phys {\bf B 383}, 643 (1992).

\bibitem{gr-zpc}  M. Gl\"{u}k, E. Reya and A. Vogt,
Z. Phys. {\bf C 53}, 127 (1992).

\bibitem{peterson} C. Peterson, D. Schlatter, J. Schmitt and
P. Zerwas, Phys. Rev. {\bf D 27}, 105 (1983).

\bibitem{dh-plb}  K.P. Das and R.C. Hwa, Phys. Lett. {\bf B 68}, 459
(1977).

\bibitem{ranft-pr} J. Ranft, Phys. Rev. {\bf D 18}, 1491 (1978).

\bibitem{nos}  E. Cuautle, G. Herrera and J. Magnin in preparation.

\bibitem{herrera}  G. Herrera and L. Montano, {\it Phys. Lett. }{\bf B 381} 
337(1996),
J. dos Anjos, G. Herrera, J. Magnin and F.R.A. Sim\~ao, in preparation.

\end{references}
\end{document}